\documentclass[aps,prc,twocolumn,groupedaddress,showpacs]{revtex4}
\usepackage{graphicx}

\begin{document}


\title{Proton-decaying states in $^{22}$Mg and the nucleosynthesis of $^{22}$Na in novae}


\author{B. Davids}
\altaffiliation{Present address: TRIUMF, 4004 Wesbrook Mall, Vancouver BC V6T2A3, Canada}
\email[]{davids@triumf.ca}
\affiliation{Kernfysisch Versneller Instituut, Zernikelaan 25, 9747 AA Groningen, The Netherlands}
\author{J. P. M. Beijers}
\author{A. M. van den Berg}
\author{P. Dendooven}
\author{S. Harmsma}
\affiliation{Kernfysisch Versneller Instituut, Zernikelaan 25, 9747 AA Groningen, The Netherlands}
\author{M. Hernanz}
\affiliation{Institut d'Estudis Espacials de Catalunya, Ed. Nexus-201, C/Gran Capit\`a 2-4, 08034 Barcelona, Spain}
\author{M. Hunyadi}
\author{M. A. de Huu}
\affiliation{Kernfysisch Versneller Instituut, Zernikelaan 25, 9747 AA Groningen, The Netherlands}
\author{J. Jos\'e}
\affiliation{Institut d'Estudis Espacials de Catalunya, Ed. Nexus-201, C/Gran Capit\`a 2-4, 08034 Barcelona, Spain}
\affiliation{Departament de F\'{\i}sica i Enginyeria Nuclear, Universitat Polit\`ecnica de Catalunya, Av. V\'{\i}ctor Balaguer s/n E-08800 Vilanova i la Geltr\'u, Barcelona, Spain}
\author{R. H. Siemssen}
\author{H. W. Wilschut}
\author{H. J. W\"{o}rtche}
\affiliation{Kernfysisch Versneller Instituut, Zernikelaan 25, 9747 AA Groningen, The Netherlands}


\date{\today}

\begin{abstract}
Populating states in $^{22}$Mg via the ($p,t$) reaction in inverse kinematics with a 55 MeV/nucleon $^{24}$Mg beam, we have measured the proton-decay branching ratios of the levels at 5.96 MeV and 6.05 MeV and obtained an experimental upper limit on the branching ratio of the 5.71 MeV state. On the basis of the present and previous measurements, we assign spins and parities to the 5.96 MeV and 6.05 MeV states. We combine our branching ratios with independent measurements of the lifetimes of these states or their $^{22}$Ne analogs to compute the resonance strengths and thereby the astrophysical rate of the $^{21}$Na($p,\gamma)^{22}$Mg reaction. We perform hydrodynamic calculations of nova outbursts with this new rate and analyze its impact on $^{22}$Na yields.
\end{abstract}


\maketitle

\def\power#1{\mbox{$\times10^{#1}\ $}}
\newcommand{\msun}{M$_\odot$ }
\newcommand{\msyr}{M$_\odot$ yr$^{-1}$ }
\newcommand{\nam}{$^{21}$Na }
\newcommand{\na}{$^{22}$Na }
\newcommand{\naa}{$^{23}$Na }
\newcommand{\netonaa}{$^{20}$Ne($p,\gamma$)$^{21}$Na($\beta^+$)$^{21}$Ne(p,
$\gamma$)$^{22}$Na($\beta^+$)$^{22}$Ne($p,\gamma$)$^{23}$Na }
\newcommand{\neo}{$^{20}$Ne }
\newcommand{\nee}{$^{21}$Ne }
\newcommand{\neee}{$^{22}$Ne }
\newcommand{\nepgnam}{$^{20}$Ne($p,\gamma$)$^{21}$Na }
\newcommand{\nepgna}{$^{21}$Ne($p,\gamma$)$^{22}$Na }
\newcommand{\nepgnaa}{$^{22}$Ne($p,\gamma$)$^{23}$Na }
\newcommand{\mgm}{$^{23}$Mg }
\newcommand{\napgmg}{$^{22}$Na($p,\gamma$)$^{23}$Mg }
\newcommand{\mgbna}{$^{23}$Mg($\beta^+$)$^{23}$Na }
\newcommand{\mgbnaa}{$^{22}$Mg($\beta^+$)$^{22}$Na }
\newcommand{\nabne}{$^{21}$Na($\beta^+$)$^{21}$Ne }
\newcommand{\nampgmg}{$^{21}$Na($p,\gamma$)$^{22}$Mg }
\newcommand{\natona}{$^{21}$Na($p,\gamma$)$^{22}$Mg($\beta^+$)$^{22}$Na }
\newcommand{\natomg}{$^{23}$Na($p,\gamma$)$^{24}$Mg }
\newcommand{\napane}{$^{23}$Na($p,\alpha$)$^{20}$Ne }
\newcommand{\mgpgal}{$^{23}$Mg($p,\gamma$)$^{24}$Al }

\section{Introduction}
The thermonuclear runaway model of classical novae \cite{starrfield72,jose98} provides a framework capable of explaining many of their features. Astronomical observations increasingly constrain current hydrodynamic simulations. One important observational test of the models would be the detection of $\gamma$-rays emitted by nuclei synthesized in a nova outburst \cite{clayton74}. Since classical novae synthesize $^{22}$Na, the 1.275 MeV $\gamma$-ray emitted by its $\beta^+$ daughter $^{22}$Ne is a good candidate for such detection efforts. An observational campaign conducted with the COMPTEL instrument on NASA's Compton Gamma Ray Observatory failed to positively detect any $^{22}$Na $\gamma$-rays from galactic novae, placing a tight upper limit on their $^{22}$Na production and ejection \cite{iyudin95}. Searches will continue with balloon- and space-borne $\gamma$-ray observatories such as the recently launched INTEGRAL of the European Space Agency.

\begin{figure}\includegraphics[width=0.5\linewidth]{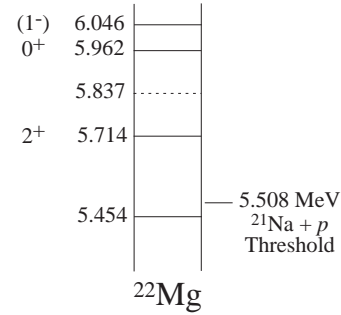} \caption{$^{22}$Mg energy level diagram near the proton threshold. Solid lines denote known states, while the dotted line indicates the position of a state whose existence is questionable. The angular momentum assignments are discussed in Section \protect\ref{rs}.} \label{fig1} \end{figure}

There are two paths by which $^{22}$Na is synthesized in novae, $^{21}$Na($\beta^+)^{21}$Ne($p,\gamma)^{22}$Na and $^{21}$Na($p,\gamma)^{22}$Mg($\beta^+)^{22}$Na. At low temperatures, the former path dominates, while at high temperatures the $^{21}$Na($p,\gamma)^{22}$Mg rate can compete with the $^{21}$Na $\beta^+$ decay rate, and the latter path becomes important. Knowledge of the $^{21}$Na($p,\gamma)^{22}$Mg rate is essential for making quantitative predictions about the total $^{22}$Na yield in a nova explosion \cite{jose99}. At low temperatures, this reaction proceeds dominantly by direct capture, but at higher temperatures, resonant contributions from excited states in $^{22}$Mg become important. Fig.\ \ref{fig1} shows a level diagram of $^{22}$Mg near the proton threshold. Recently, the first direct measurement of the resonance strength of the 5.714 MeV state in $^{22}$Mg that dominates the $^{21}$Na($p,\gamma)^{22}$Mg reaction rate at nova temperatures was carried out at TRIUMF \cite{bishop03}.

In this paper, we describe a measurement of proton-decay branching ratios of states in $^{22}$Mg lying just above the proton separation energy. We obtain an upper limit on the proton-decay branching ratio of the 5.714 MeV state. Combining this upper limit with an independent measurement of the lifetime of this state, we calculate an upper limit on its resonance strength that is consistent with the measurement of Ref.\ \cite{bishop03}. We use the results of these two measurements to determine the most likely value of this resonance strength. Taking into consideration the experimental information on higher-lying states in $^{22}$Mg that may also contribute to the $^{21}$Na($p,\gamma)^{22}$Mg rate in novae, we calculate the astrophysical reaction rate and perform hydrodynamic calculations of nova outbursts to study its impact on $^{22}$Na yields.

\section{Experimental Method}
The detailed balance theorem implies that the same information can be gained from studying the decay of a resonance as can be learned from measuring its formation. Therefore we have performed a measurement of the proton-decay branching ratios of the states in $^{22}$Mg relevant to the $^{21}$Na($p,\gamma)^{22}$Mg reaction rate in novae. The measurement was carried out at the Kernfysisch Versneller Instituut (KVI) using a recoil coincidence technique \cite{davids03} through which we have detected both proton- and $\gamma$-decaying recoils with 100\% geometric efficiency in the Big-Bite Spectrometer \cite{berg95}. A 55 MeV/nucleon $^{24}$Mg beam provided by the variable energy, superconducting cyclotron AGOR bombarded a 1 mg cm$^{-2}$ (CH$_2)_n$ target to populate states in $^{22}$Mg via the ($p,t$) reaction in inverse kinematics.  Both triton ejectiles and $^{22}$Mg recoils entered the magnetic spectrometer, which was positioned at 0$^{\circ}$. The $^{22}$Mg recoils with excitation energies less than 8 MeV deexcited by the emission of $\gamma$ rays, retaining their identities as $^{22}$Mg, or by the emission of protons, resulting in the formation of $^{21}$Na decay products. Proton decays were identified through $^{21}$Na-triton coincidences, while $\gamma$ decays were observed as $^{22}$Mg-triton coincidences. 

Heavy recoils and decay products were detected in two phoswich detectors \cite{leegte92}, while two vertical drift chambers \cite{woertche01} recorded the positions and angles of the triton ejectiles before they were stopped in a separate array of six phoswiches. Particles were identified using energy loss, total energy, and time-of-flight information provided by the phoswich detectors. The experimental technique and apparatus were previously employed in our measurement of $^{19}$Ne $\alpha$-decay branching ratios, and we refer to Ref.\ \cite{davids03a} for a complete description.

Strong kinematic forward focusing in the bombardment of a proton target by 1.3 GeV $^{24}$Mg projectiles made possible full angular and momentum acceptance for recoils and decay products. There are two solutions of the reaction kinematics at laboratory angles around 0$^{\circ}$, one in which tritons are emitted forward in the center-of-mass system, and one in which they are emitted backward. We detected those emitted backward in the center-of-mass system, which have laboratory energies around 19 MeV/nucleon. For triton ejectiles detected at laboratory scattering angles of 4$^{\circ}$ or less in this measurement, the $^{22}$Mg recoils emerged at scattering angles up to 0.3$^{\circ}$. The impulse delivered to the $^{21}$Na decay product in a proton decay results in a small angular spread about the original $^{22}$Mg trajectory, but the high incident beam energy and low decay energies of the states studied limited the laboratory scattering angles of the $^{21}$Na decay products to 0.5$^{\circ}$. Simultaneous detection of both ejectile and recoil or decay product is possible because the large momentum acceptance of the spectrometer can accomodate the 12\% magnetic rigidity difference between $^{21}$Na decay products and triton ejectiles. 

\section{Experimental Results}
\begin{figure}\includegraphics[width=\linewidth]{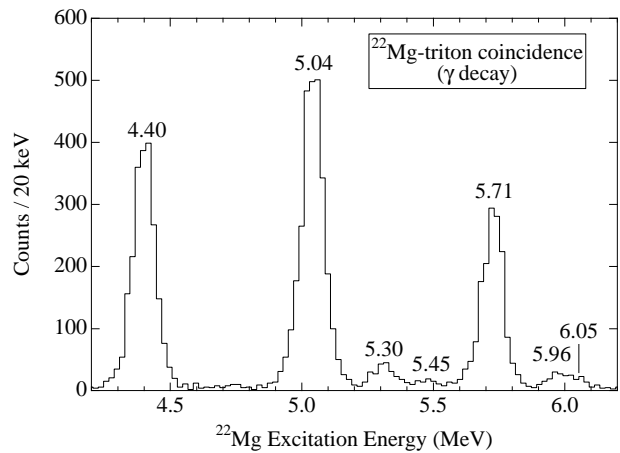} \caption{$^{22}$Mg excitation energy spectrum obtained from $^{22}$Mg-triton coincidences, representing $\gamma$ decays of states in $^{22}$Mg. Known states are labeled by their excitation energies.} \label{fig2} \end{figure}

\begin{figure}\includegraphics[width=\linewidth]{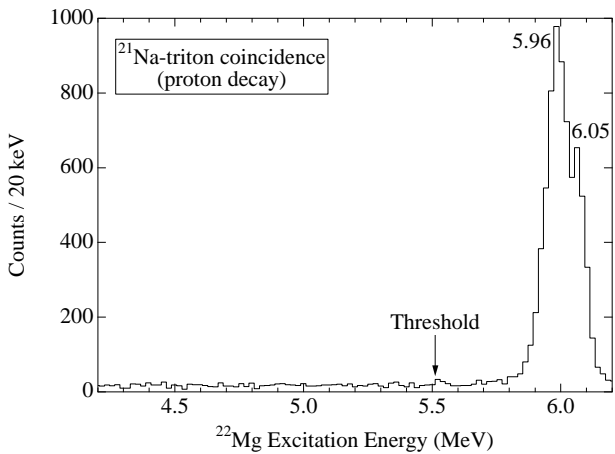} \caption{$^{22}$Mg excitation energy spectrum obtained from $^{21}$Na-triton coincidences, representing proton decays of states in $^{22}$Mg. The proton-decay threshold lies at 5.5 MeV.} \label{fig3} \end{figure}

Excitation energies of the $^{22}$Mg recoils were reconstructed from the measured momenta of the triton ejectiles. The $\gamma$-decay spectrum obtained from $^{22}$Mg-triton coincidences is shown in Fig.\ \ref{fig2}, and the proton-decay spectrum obtained from $^{21}$Na-triton coincidences is displayed in Fig.\ \ref{fig3}. States in $^{22}$Mg are labeled by their excitation energies. Several recent experiments have studied the states near the proton threshold \cite{bateman01,chen01,michimasa02, caggiano02,berg03}, and the experimental developments since the last evaluation \cite{endt90} are nicely summarized in Ref.\ \cite{caggiano02}. Comparing the excitation energies of the seven states observed in this measurement with the literature values \cite{caggiano02}, we find a root mean square deviation of 6 keV. An excitation energy resolution of  90 keV full-width-at-half-maximum was obtained. This resolution is insufficient to completely separate the 5.96 and 6.05 MeV states, so a fit consisting of Gaussians and a constant background was used to determine the yield to these two states in the $\gamma$-decay and proton-decay spectra.

The proton separation energy in $^{22}$Mg is 5.5 MeV \cite{audi95}, so of the states observed here, only those lying at 5.71, 5.96, and 6.05 MeV can decay by proton emission. While proton decay is the dominant deexcitation mechanism for the two states around 6 MeV, Fig.\ \ref{fig3} reveals no statistically significant evidence for proton decays of the 5.714 MeV state. Hence we set an upper limit on the proton decay yield from this state using Bayesian statistics with a uniform prior probability density function \cite{hagiwara02}. As the thresholds for the emission of neutrons and $\alpha$ particles lie at higher excitation energies, only proton and $\gamma$ decay are possible for these states. Therefore, the proton-decay branching ratio is given by B$_p\equiv\Gamma_p/\Gamma=\Gamma_p/(\Gamma_p+\Gamma_{\gamma})$. The proton-decay branching ratios deduced from this measurement are given in Table \ref{table1}.

\begin{table}
 \caption{\label{table1}Proton-decay branching ratios B$_p$, decay widths, and resonance strengths of states in $^{22}$Mg. Unless otherwise noted, upper limits  and uncertainties are specified at the 95.5\% confidence level.}
 \begin{ruledtabular}
 \begin{tabular}{cccccc}
E (MeV)&J$^{\pi}$&B$_{p}$&$\Gamma_{\gamma}$ (meV)&$\Gamma$ (meV)&$\omega\gamma$ (meV)\\
\hline
5.714&$2^+$&$\leq 0.020$&&$16 ^{+50}_{-7}$\footnote{Ref.\ \cite{grawe75}}&$\leq 0.80$\\
5.962&$0^+$&$0.98(1)$&$2 ^{+5}_{-1}$\footnote{Value from $^{22}$Ne analog state \cite{endt98}}&&$0.2 ^{+0.6}_{-0.1}$\\
6.046&(1$^-$)\footnote{See the text for a discussion of this tentative assignment.}&$0.97(3)$&$2 ^{+2}_{-1}$\footnote{Value from $^{22}$Ne analog state \cite{olivier93}. Uncertainties in the width and resonance strength are 1$\sigma$.}&&$0.7 ^{+0.8}_{-0.3}$\\
\end{tabular}
\end{ruledtabular}
\end{table}

\section{Resonance Strengths}\label{rs}

Of the three proton-decaying states observed here, $\gamma$-decays of only the 5.714 MeV state have been studied previously. Its mean lifetime has been measured to be $40\pm15$ fs \cite{grawe75}. The corresponding total decay width is given in Table \ref{table1}. In the case of the other two states, no experimental information on the radiative or total decay widths is known. For the purpose of estimating their resonance strengths we have assumed that they have the same radiative widths as their analog states in $^{22}$Ne \cite{olivier93,endt98}.

Definitive angular momentum and parity assignments for the two higher-lying states have proven somewhat difficult to make. In fact, the existence of two distinct states around 6 MeV only became clear as a result of the ($p,t$) measurement of Ref.\ \cite{bateman01}, in which both  were observed simultaneously. The latest compilation \cite{endt90} includes only a single level at 5965(25) keV. The angular momentum and parity of this state were assigned as 0$^+$ on the basis of angular distribution measurements using the ($^3$He,$n$) reaction \cite{mcdonald70,alford86}. In an early ($p,t$) measurement at 42 MeV \cite{paddock72}, a peak was seen at 6.061(37) MeV that was assumed to be the same 0$^+$ state observed in the ($^3$He,$n$) study at 5.945(20) MeV. Its angular distribution was compared with DWBA calculations and found to be consistent with a $0^+$ assignment. 

In the recent ($p,t$) measurements at 38 MeV \cite{bateman01} and 35 MeV \cite{michimasa02}, the 6.046 MeV state was found to be populated about ten times more strongly than the 5.962 MeV state. Those measurements covered center-of-mass angles larger than about 10$^{\circ}$. On the basis of angular distribution measurements and comparison with the analog $^{22}$Ne nucleus, the authors of Ref.\ \cite{michimasa02} assert that the 6.046 MeV state is 0$^+$ and tentatively assign 1$^-$ to the 5.962 MeV level. However, their calculated $L=0$ and $L=1$ angular distributions appear to describe both transitions equally well. Most importantly, the angular distributions of Refs.\ \cite{paddock72,michimasa02} do not cover the small angles at which $L=0$ transitions are most readily identified. In contrast, the present ($p,t$) data taken at 0$^{\circ}$ indicate that the 5.962 MeV state is approximately four times more strongly populated than the 6.046 MeV level at forward angles. On the basis of its strong peaking at 0$^{\circ}$ and the ($^3$He,$n$) data, we conclude contrary to Ref.\ \cite{michimasa02} that the spin and parity of the 5.962 MeV state is 0$^+$. This conclusion is corroborated by DWBA calculations we carried out that consistently describe the beam-energy dependence of the relative intensities and angular distributions of the 5.962 MeV and 6.046 MeV states with $L=0$ and $L=1$ transitions respectively. While we regard the 0$^+$ assignment as definitive, the 1$^-$ assignment for the 6.046 MeV level must remain tentative.

Given the angular momentum J, proton width $\Gamma_p$, radiative width $\Gamma_{\gamma}$, and total width $\Gamma$ of a resonance, its strength is given by \begin{equation}\omega\gamma\equiv\frac{2\mathrm{J} +1}{(2\mathrm{J_{^{21}Na}}+1)(2\mathrm{J}_{p}+1)}\frac{\Gamma_{p}\Gamma_{\gamma}}{\Gamma},\end{equation} which reduces to \begin{equation}\omega\gamma=\frac{2\mathrm{J} +1}{8}B_{p}(1-B_p)\Gamma=\frac{2\mathrm{J} +1}{8}B_{p}\Gamma_{\gamma}.\end{equation} We use this expression to compute the resonance strength of the 5.962 MeV level, and, using a $2\sigma$ upper limit on $\Gamma$ of 65.8 meV, to calculate a 95.5\% confidence level upper limit on the resonance strength of the 5.714 MeV level. These values are shown in Table \ref{table1}. Our upper limit is consistent with the measurement of Ref.\ \cite{bishop03}, $1.03\pm0.16$ (stat.) $\pm$ 0.14 (sys.) meV. Adding the systematic and statistical uncertainties of this measurement in quadrature, we combine it with the present result and find a joint 95.5\% confidence level interval of 0.60-0.80 meV.  

We now address the issue of the resonance energies. Although the excitation energies of the levels with respect to the ground state of $^{22}$Mg are well determined, the proton separation energy is less well defined. A recent direct measurement of the 5.714 MeV state with a $^{21}$Na beam at TRIUMF \cite{bishop03} found the energy of this resonance to be 205.7(5) keV, rather than 212 keV, which is the difference between the measured excitation energy of this level \cite{grawe75} and the proton separation energy given in Ref.\ \cite{audi95}. This implies that the mass excess of $^{22}$Mg is some 6 keV smaller than the published value \cite{audi95}. Such an interpretation is consistent with a recent reanalysis of one of the experiments used to deduce the $^{22}$Mg mass excess \cite{hardy03}. Hence we adjust the separation energy and take the resonance energies to be 206 keV and 454 keV for the 5.714 MeV and 5.962 MeV levels, respectively.

The most recent compilation of levels in $^{22}$Mg \cite{endt90} lists a state at 5.837 MeV. This level was identified on the basis of a single ($^3$He,$n\gamma$) measurement \cite{rolfs72}, but has not been seen in any other experiment, including an independent ($^3$He,$n\gamma$) measurement \cite{grawe75}, two ($^3$He,$n$) measurements \cite{mcdonald70,alford86}, three ($p,t$) measurements \cite{paddock72,bateman01,michimasa02}, an ($^{16}$O,$^6$He) measurement \cite{chen01}, an ($^3$He,$^6$He) measurement \cite{caggiano02}, an ($^4$He,$^6$He) measurement \cite{berg03}, and the present ($p,t$) measurement. Although the observation of peaks in triple coincidence spectra such as those shown in Ref.\ \cite{rolfs72} is usually taken as sufficient evidence for transitions, the presence of contaminant lines in these spectra implies that the $\gamma$-ray transitions attributed to a 5.837 MeV level may be due to target impurities. In our view, the weight of the available experimental evidence is insufficient to support the existence of a level at 5.837 MeV in $^{22}$Mg.

\section{Astrophysical Reaction Rate}

\begin{figure}\includegraphics[width=\linewidth]{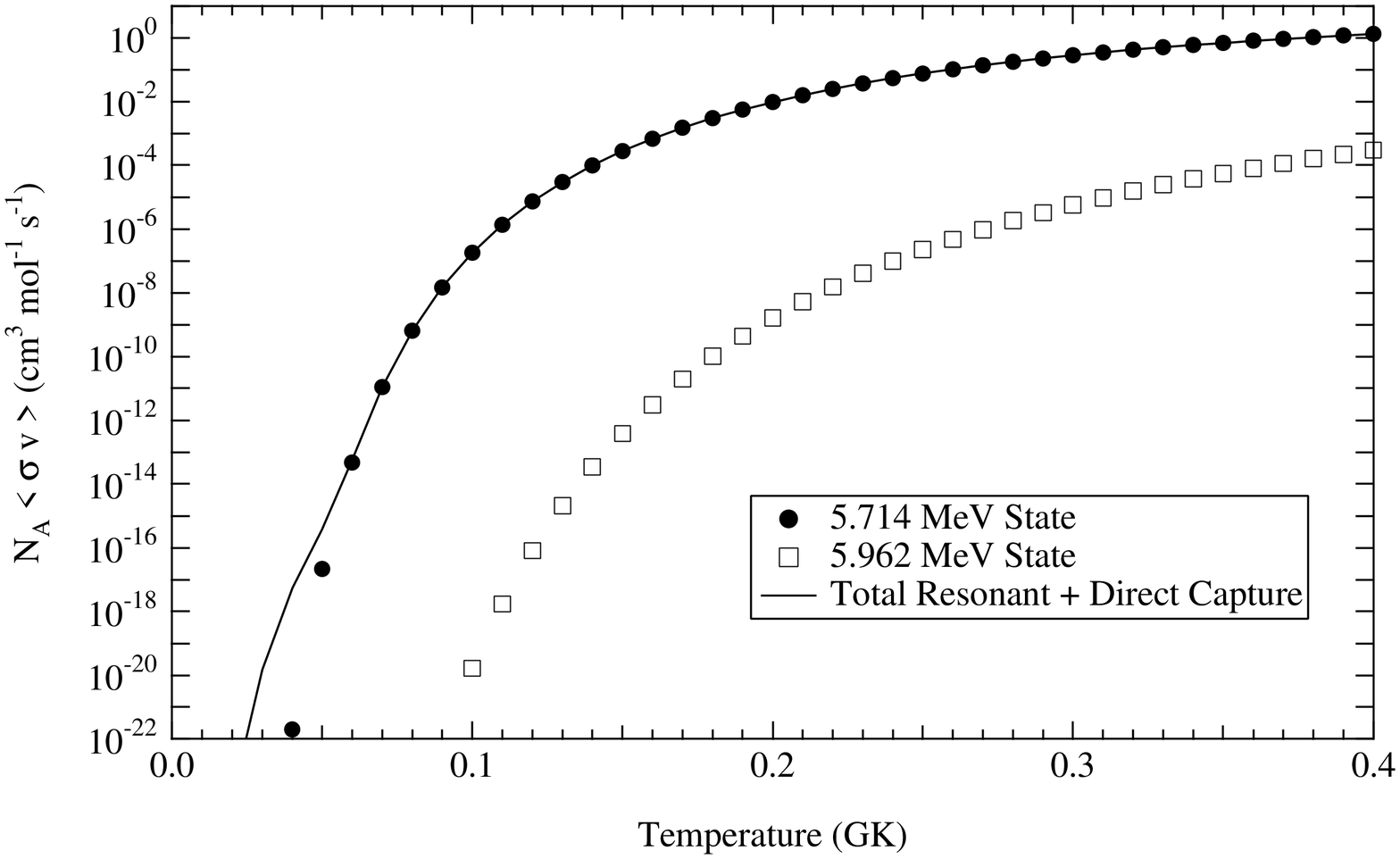} \caption{Product of the Avogadro constant N$_{\mathrm{A}}$ and the thermally averaged rate of the $^{21}$Na$(p,\gamma)^{22}$Mg reaction per particle pair. Contributions from the two most important resonances are shown, along with the sum of the resonant and direct capture rates.} \label{fig4} \end{figure}

We have calculated the thermally averaged $^{21}$Na$(p,\gamma)^{22}$Mg reaction rate per particle pair as a function of temperature, including direct capture and the resonant contributions of the states at 5.714 and 5.962 MeV. Fig.\ \ref{fig4} shows the individual contributions of the two most important resonances and the sum of the resonant and direct capture rates. A resonance strength of 0.8 meV is used for the 5.714 MeV state, as this is the most likely value within the joint 95.5\% confidence level interval of the TRIUMF \cite{bishop03} and present measurements. It can be seen that the 5.96 MeV level does not contribute significantly to the reaction rate at temperatures characteristic of novae. Only direct capture and the resonant contribution of the 5.71 MeV state are important in this regime.

The direct capture contribution was calculated using the low-energy S factor given in Ref.\ \cite{bateman01}, which has a value of 7.9 keV b at zero energy. This S factor is roughly 40\% larger than that calculated in Ref.\ \cite{wiescher86}. An apparent discrepancy between the direct capture contributions calculated using these two S factors, shown in Table XI of Ref.\ \cite{bateman01}, is due to an erroneous parametrization of the direct capture contribution taken from Ref.\ \cite{caughlan88} rather than to the difference between the calculated S factors. Direct capture only contributes significantly to the reaction rate below 0.07 GK.

In Ref.\ \cite{bishop03}, the total $^{21}$Na$(p,\gamma)^{22}$Mg reaction rate is calculated using the same analytic fit given in Ref.\ \cite{jose99}, but taking the coefficients for the term due to the 5.714 MeV state from the direct measurement of the strength of this resonance. The reaction rate calculated here differs from that used in Ref.\ \cite{bishop03} in several respects. First, the direct capture contribution here is several orders of magnitude larger at the lowest temperatures due to the parametrization error mentioned above. Second, the contribution of the putative 5.837 MeV state included in the previous rate is excluded here due to our doubts over the existence of this state. Third, the contribution of the 5.962 MeV state is about ten times smaller here than in the previous rate, for which the resonance strength was assumed to be 2.5 meV. Finally, the resonance strength of the 5.714 MeV state adopted here is about 20\% smaller than that measured in Ref.\ \cite{bishop03}. The overall effect of all of these changes is to reduce the total reaction rate to approximately 70\% of the rate employed in Ref.\ \cite{bishop03} in the temperature range from 0.1-0.33 GK important for $^{22}$Na synthesis in novae. 

\section{Synthesis of $^{22}$Na in Classical Novae}

\begin{figure*}\includegraphics[width=\linewidth]{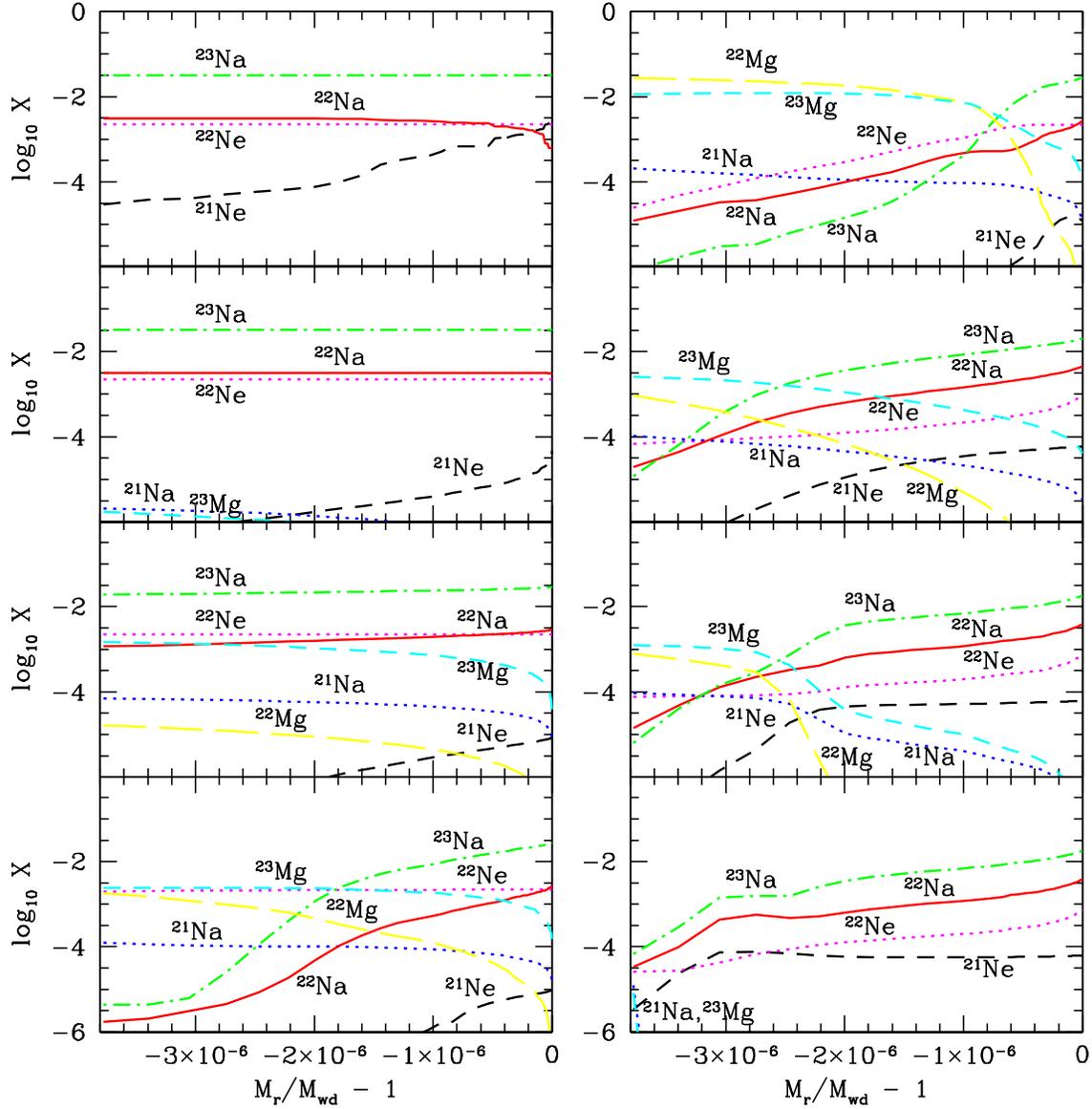} \caption{(Color online) Mass fractions of various isotopes in logarithmic scale along the accreted envelope for a 1.35 solar mass ONe nova accreting at a rate $\dot M = 2 \times 10^{-10}$ $M_\odot$ yr$^{-1}$. The mass coordinate represents the mass below the surface (M$_r$) relative to the total white dwarf mass (M$_{wd}$). Thus the left side of each panel corresponds to the base of the accreted envelope, and the right side its outer surface. From top to bottom and then left to right, the first five panels show a time series from the early stages of the explosion up to the ejection stage, with a temperature in the burning shell equal to: 0.07, 0.09, 0.15, 0.25, and $T_{peak} = 0.327$ GK. The last three panels correspond to the final phases of the explosion, when the white dwarf envelope has expanded to a size of $R_{wd} \sim 10^9, 10^{10}$, and $10^{12}$ cm, respectively. All of the material lying at a mass coordinate $M_r/M_{wd}-1\geq-3.05\times10^{-6}$ is ejected by the explosion.} \label{fig5} \end{figure*}

\begin{figure*}\includegraphics[width=\linewidth]{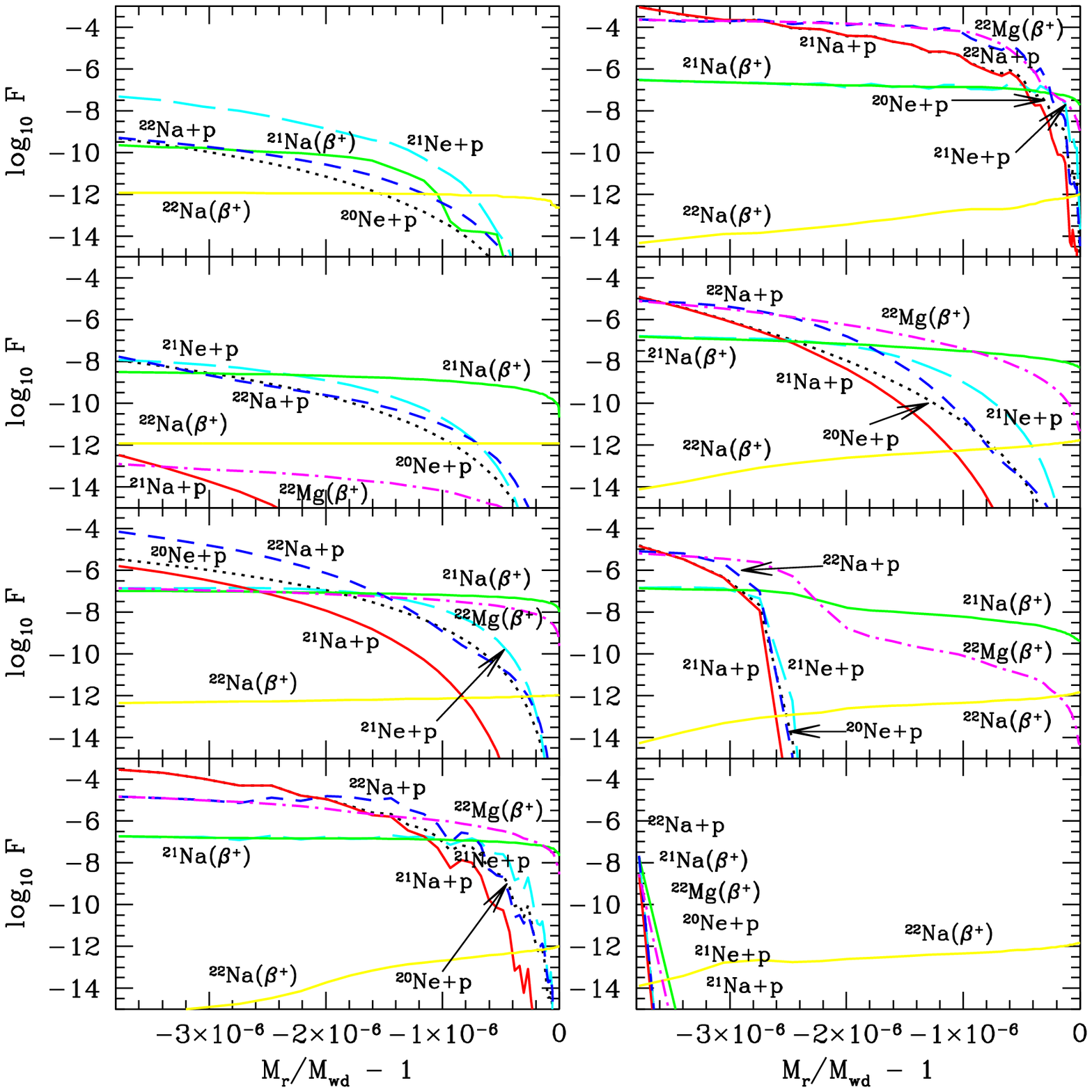} \caption{(Color online) Same as the previous figure, but plotting reaction and decay rates per unit volume (cm$^{-3}$ s$^{-1}$).} \label{fig6} \end{figure*}

The high peak temperatures attained in Oxygen-Neon (ONe) nova outbursts of $\sim 2 - 3.5 \times 10^8$ K allow for nuclear activity in the NeNa and MgAl cycles, resulting in significant production of astrophysically interesting species such as $^{22}$Na and $^{26}$Al. Here we describe the main path leading to $^{22}$Na synthesis in novae. In order to check the influence of our new $^{21}$Na($p,\gamma$) rate, we have performed two different hydrodynamic calculations of nova outbursts on 1.25 and 1.35 \msun ONe white dwarfs accreting solar-like matter at a rate $\dot M = 2 \times 10^{-10}$ \msyr. Our models assume 50\% mixing between the solar-like accreted envelope and the outermost shells of the ONe core \cite{jose98}. The nuclear reaction network includes the new $^{21}$Na($p,\gamma$) rate discussed in this paper. Snapshots of the abundance evolution of several isotopes relevant to $^{22}$Na synthesis during a nova outburst on a 1.35 \msun white dwarf, $^{21,22}$Ne, $^{21,22,23}$Na and $^{22,23}$Mg, are shown in Fig.\ \ref{fig5}. Reaction and decay rates per unit volume from the same simulation are shown in Fig.\ \ref{fig6} for a number of relevant nuclear reactions and $\beta$ decays, e.g., $^{20}$Ne($p,\gamma$), $^{21}$Ne($p,\gamma$), $^{21}$Na($p,\gamma$), $^{21}$Na($\beta^+$), $^{22}$Na($p,\gamma$) $^{22}$Na($\beta^+$), and $^{22}$Mg($\beta^+$).
 
During the early stages of the outburst at the onset of accretion, the evolution of $^{22}$Na is dominated by the cold NeNa cycle, which proceeds via $^{20}$Ne($p,\gamma$)$^{21}$Na($\beta^+$)$^{21}$Ne($p,\gamma$)$^{22}$Na. As soon as a critical mass accumulates on the white dwarf, a thermonuclear runaway ensues, leading to a sudden temperature rise.  When the temperature at the burning shell reaches $T_{bs} =$ 0.07 GK, the main nuclear reaction of the NeNa cycle is $^{21}$Ne($p,\gamma$)$^{22}$Na, which essentially decreases the abundance of $^{21}$Ne. This reduction becomes increasingly important as the temperature rises, as can be seen in panel 1 of Fig.\ \ref{fig5}, which corresponds to a temperature of $T_{bs}$ = 0.07 GK. At this stage there is not enough $^{21}$Ne to feed the main chai§n of reactions leading to $^{22}$Na synthesis through $^{21}$Ne($p,\gamma$). Therefore, because of efficient proton captures in the inner part of the envelope, the abundance of $^{22}$Na decreases with the rise of temperature towards the peak value (c.f. Fig.\ \ref{fig5}, panels 2, 3, \& 4). 

At $T_{bs} =$ 0.15 GK (Fig.\ \ref{fig5}, panel 3), the mass fraction of $^{21}$Ne has decreased below $10^{-6}$, except in the outer envelope where some \nee is produced via $\beta^+$ decay of $^{21}$Na. Moreover, $^{21}$Na  and $^{22,23}$Mg  increase due to $^{20}$Ne($p,\gamma$)$^{21}$Na and $^{21,22}$Na($p,\gamma$)$^{22,23}$Mg, respectively.
As shown in panel 3 of Fig. 6, $^{21}$Na($p,\gamma$) becomes faster than $^{21}$Na($\beta^+$)
near the burning shell at this stage, which marks the onset of the hot NeNa cycle.
Destruction of $^{22}$Na through ($p,\gamma$) is already noticeable in panel 3, but due to
convection, $^{22}$Na shows a nearly flat profile throughout the envelope. 
It is also noteworthy that at this time $^{22}$Na($p,\gamma$) becomes the most
important reaction in the NeNa region (Fig.\ \ref{fig6}, panel 3).

When $T_{bs} =$ 0.25 GK (Fig.\ \ref{fig5}, panel 4), there is a dramatic decline in 
several isotopes, including  \naa due to ($p,\gamma$) and ($p,\alpha$) reactions, and  \na due to
\napgmg.  The abundance of \nam is only slightly larger than it was at 0.15 GK (panel 3) because of a quasi-equilibrium between \nampgmg and \nepgnam near the envelope's base (Fig.\ \ref{fig6}, panel 4). Mass fractions of $^{22,23}$Mg increase due to proton captures on $^{21,22}$Na respectively.

Seven seconds later, the temperature in the burning shell attains its peak value of 0.327 GK (Fig.\ \ref{fig5}, panel 5). The mean amount of \na increases due to a contribution from $^{22}$Mg($\beta^+$) previously transported by convection to the outer, cooler layers of the envelope. In fact, both $^{22,23}$Mg attain maximal abundance at this stage as a result of proton captures on $^{21,22}$Na, with $^{22}$Mg being the most abundant isotope of this mass region. Some increase in the $^{21}$Na abundance results from the fact that whereas in the inner part of the envelope \nampgmg and \nepgnam are still in quasi-equilibrium, the outer envelope is clearly dominated by $^{20}$Ne($p,\gamma$); $^{21}$Na nuclei are rapidly transported inwards by means of convective motions. The amount of \naa decreases due to both ($p,\gamma$) and ($p,\alpha$) reactions, which dominate \nepgnaa as well as \mgbna. Meanwhile, \neee is efficiently destroyed by \nepgnaa. 

 Shortly after the peak temperature is reached, the envelope begins to expand due to the sudden release of energy from the short-lived $\beta^+$ emitters $^{13}$N, $^{14,15}$O, and $^{17}$F. The role played by ($p,\gamma$) and ($p,\alpha$) reactions is therefore reduced following the drop in temperature, and $\beta^+$ decays progressively dominate the evolution (Fig.\ \ref{fig6}, panels 6 \& 7). The abundances of $^{21}$Na and $^{22,23}$Mg decrease as a result of these $\beta^+$ decays, which in turn increase the amount of $^{21}$Ne and $^{22,23}$Na (Fig.\ \ref{fig5}, panels 6 \& 7).

At the final stages of the outburst (Figs. 5 \& 6, panel 8),  as the envelope expands and cools, most of the remaining nuclear activity in the NeNa cycle is due to $\beta^+$ decays, in particular $^{22}$Na($\beta^+$), since other short-lived species such as $^{21}$Na and $^{22,23}$Mg have already decayed. The resulting mean mass fraction of \na in the ejected shells in this $1.35$ \msun white dwarf model is X(\na) = 1.1 \power{-3}, which corresponds to 4.7\power{-9} \msun of \na ejected into the interstellar medium. Other species of the NeNa cycle present in the ejecta are \neo with X = 0.14, scarcely modified with respect to the initial value of 0.16 because of the moderate peak temperature achieved, $^{21,22}$Ne with X = 6.2\power{-5} and 1.8\power{-4} respectively, and \naa with X = 5.6\power{-3}.

The present 1.35 \msun ONe white dwarf simulation yields an ejected $^{22}$Na mass fraction very similar to the previous estimate of $1 \times 10^{-3}$ \cite{jose99}, reflecting a combination of two different effects: an increase of the $^{21}$Na($p,\gamma$) rate associated with the new determination of the resonance strength of the 5.714 MeV level that tends to reduce the final $^{22}$Na yield, and the elimination of the 5.837 MeV level that reduces the rate, thereby increasing the final $^{22}$Na abundance. The peak temperatures reached in the 1.35 \msun ONe white dwarf outburst would imply a greater role for the putative 5.837 MeV level in the overall rate compared to the 1.25 \msun simulation, which reaches a lower peak temperature of 0.25 GK. A mean $^{22}$Na mass fraction of $2.8 \times 10^{-4}$, averaged within the ejected envelope shells, is obtained for the 1.25 \msun ONe white dwarf nova simulation. This is to be compared with the value of $3.5\times10^{-4}$ obtained in Ref.\ \cite{jose99}. The small reduction in the final mass fraction is due to the larger resonance strength adopted for the 5.714 MeV level here. Our results for the 1.25 \msun model are indistinguishable from those obtained with the recent $^{21}$Na($p,\gamma$) rate based on the first direct measurement of the resonance strength of the 5.714 MeV level \cite{bishop03}, illustrating the modest effect a level at 5.837 MeV would have on the overall rate in such explosions.

\section{Summary}

We have carried out a measurement of the proton-decay branching ratios of near-threshold states in $^{22}$Mg. Populating the states via the ($p,t$) reaction in inverse kinematics with a 55 MeV/nucleon $^{24}$Mg beam, we have detected tritons in coincidence with $^{22}$Mg recoils and $^{21}$Na decay products with 100\% geometric efficiency. On the basis of the present and previous measurements, we assign spins and parities to the 5.96 MeV and 6.05 MeV levels. We combine our branching ratios with independent measurements of the lifetimes of the states and compute their resonance strengths. Calculating the astrophysical rate of the $^{21}$Na($p,\gamma)^{22}$Mg reaction based on all available experimental information, we have performed hydrodynamic calculations of nova outbursts and analyzed the impact of this new rate on $^{22}$Na yields. Comparison with a previous calculation \cite{jose99} reveals that this new rate changes the predicted amount of ejected $^{22}$Na only slightly. 

\begin{acknowledgments}
This work was performed as part of the research program of the {\it Stichting voor Fundamenteel Onderzoek der Materie} with financial support from the {\it Nederlandse Organisatie voor Wetenschappelijk Onderzoek}.  JJ and MH acknowledge support from MCYT grants AYA2001-2360, AYA2002-04094-C03-02 and AYA2002-04094-C03-03.
\end{acknowledgments}
\bibliography{s27}
\end{document}